\documentclass[letter]{aa} % for the letters 
\usepackage{graphicx}
\usepackage{txfonts}
\begin{document}

        \title{Anti-correlation between multiplicity and orbital properties in exoplanetary systems as a possible record of their  dynamical histories}
        
        \titlerunning{Exoplanetary systems multiplicity trends}
        
        %\subtitle{I. Overviewing the $\kappa$-mechanism}
        
        \author{A. Zinzi
                \inst{1}\fnmsep\inst{2}
                \and
                D. Turrini\inst{3}\fnmsep\inst{4}
        }
        
        \institute{Space Science Data Center (SSDC) -- ASI, Via del Politecnico snc, 00133, Rome, Italy
                \and
                INAF-OAR, Via Frascati n. 33, 00078, Monte Porzio Catone (RM), Italy\\
                \email{angelo.zinzi@ssdc.asi.it}
                \and
                INAF-IAPS, Via del Fosso del Cavaliere n. 100, 00133, Rome, Italy
                \and
                Departamento de Fisica, Universidad de Atacama, Copayapu 485, Copiap{\'o}, Chile\\
                \email{diego.turrini@iaps.inaf.it}
        }
        
        \date{Received July 19, 2017; accepted August 26, 2017}
        
        % \abstract{}{}{}{}{} 
        % 5 {} token are mandatory
        
        \abstract
        {Previous works focused on exoplanets discovered with the radial velocity (RV) method reported an anti-correlation between the orbital eccentricities of the exoplanets and the multiplicity $M$ (i.e., the number of planets) of the systems they inhabit.}
        {We further investigate this reported anti-correlation here using a dataset comprising exoplanets discovered with both the RV and transit methods,  searching for hints of its causes by exploring the connection between the number of planets and the dynamical state of the exosystems.}
        {In order to examine the correlation between multiplicity and orbital eccentricity, for every multiplicity case considered ($2 \leq M \leq 6$), we computed the weighted average eccentricities instead of the median eccentricities used previously. The average eccentricities were calculated using the inverse of the uncertainty on the eccentricity values as weights. The analysis of the dynamic state of the exosystems was performed by computing their angular momentum deficit (AMD), which is a diagnostic parameter successfully used in the study of the solar system and recently applied to exosystems as well.}
        {Our results confirm the reported multiplicity-eccentricity anti-correlation and show that the use of the uncertainties on the orbital eccentricities in the analysis allows for a better agreement between the data and the fits. Specifically, our best fit reproduces well the behaviour of the average eccentricities for all systems with $M>1$, including the additional cases of TRAPPIST-1 ($M=7$) and of the solar system ($M=8$).  The AMD analysis, while not conclusive due to the limited number of exosystems that could be analysed, also suggests the existence of an anti-correlation between the multiplicity and the AMD of exosystems. This second anti-correlation, if confirmed by future studies, raises the possibility that the population of low-multiplicity exosystems is contaminated by former high-multiplicity systems that became dynamically unstable and lost some of their planets.
        }
        {}
        
        
        \maketitle
        %
        %-------------------------------------------------------------------
        
        \section{Introduction}

The currently known population of exoplanets shows characteristics that are extremely different from those of their counterparts orbiting the Sun and no exact analogue of the solar system has yet been found (see e.g. \citealt{hatzes2016} for a recent review). Albeit it is likely that our picture of exoplanetary systems could be incomplete because of the limitations of our present observational capabilities, this nevertheless reveals the large variety of exosystems across the Galaxy and the diversity of the outcomes of the processes shaping the formation and evolution of planetary systems (see \citealt{hatzes2016} and references therein).
        
Exploiting the possibility to perform statistical studies for the first time thanks to the large number of discovered exoplanets, recent works \citep{juric,limbach} reported the existence of an anti-correlation between the multiplicity of exosystems (i.e. the number of planets they possess) and the orbital eccentricities of their planets. In particular, \cite{limbach} analysed a sample of 403 exoplanets showing that, once divided in bins according to the multiplicity of their host systems, the medians of their eccentricities can be fitted by a power law. These authors also verified that the same power law can be used to fit the median of the eccentricities of solar system planets.

In building their sample, \cite{limbach} selected exoplanets exclusively discovered by means of the radial velocity (RV) method, requiring eccentricity $e$ measurements that are different from zero for all of the planets. This selection criterion ended up in a dataset comprising 276, 81, 25, 12 and 9 exoplanets for multiplicities $M$ equal to 1, 2, 3, 4 and 5-6, respectively, the last bin comprising two multiplicity values to increase the statistics. Limbach \&\ Turner performed the analysis without considering the uncertainty on eccentricity measurements and calculated a power law fit for the data with M > 1; they obtained the relation $e(M) = 0.584 M^{-1.20}$, which they found to reproduce well the median eccentricities for M > 2.       
        
In the present work we further investigated the findings of \cite{limbach} by taking into account the uncertainties on the eccentricity measurements in our analysis. To gain a deeper insight on the causes of the observed anti-correlation, we also performed a preliminary analysis of the dynamical state of the exosystems (instead of the individual exoplanets) in our sample by evaluating their angular momentum deficits \citep{AMD}.\\
This work is organized as follows: Sect. 2 is dedicated to the description and reduction of the dataset used; in Sect. 3 these data are discussed; and finally, in Sect. 4 conclusions are given.

        %--------------------------------------------------------------------
        \section{Dataset and methods}

        To compose our sample we selected planets around stars with effective temperatures between 7920 K and 2600 K (i.e. as expected for type F, G, K, and M stars) in systems with at least two planets. We used the NASA Exoplanet Archive database \citep{caltech}, which we queried by means of the Exoplanetary Analysis and 3D visualization Tool (ExoplAn3T). This web tool, currently under development at the Space Science Data Center (SSDC), allows users to select between the majority of the parameters provided by the NASA Exoplanet Archive application programming interfaces. However, since the query result also includes the required characteristics for all the objects (i.e. star and planets) that constitute the systems, the output of the tool can be used to perform analyses of the systems as a whole.
                
In our query we made no distinction on the discovery technique of the exoplanets, with the requirement of knowing the estimated uncertainties on their orbital eccentricities. The resulting dataset consisted of 258 exoplanets. Of these, 227 were detected by means of the RV method and 56 were detected with the transit method; the total number is larger than our sample since some exoplanets were observed with multiple techniques. While our sample is smaller than that used by \cite{limbach}, our selection criterion prevents us from introducing further observational biases in the analysis, allowing us to weight the data according to their reliability and information content.

\begin{figure}
                \centering
                \includegraphics[width=\hsize]{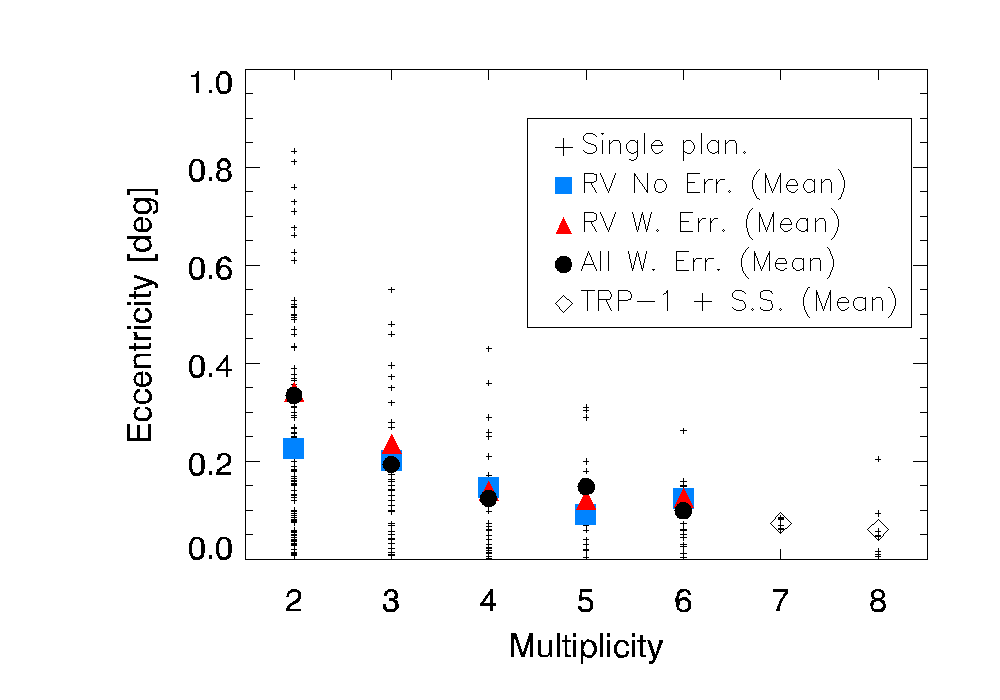}
                \caption{Multiplicity $M$ vs. eccentricity $e$ for all the planets in our sample. The cases for $M$ = 7 and $M$ = 8 are the TRAPPIST-1 planets and the planets in the solar system. The individual exoplanets in our sample are represented by black crosses. The blue squares represent the mean for every bin for planets observed with the RV method without considering errors. The red triangles are the means associated with planets detected with the RV method taking into account the uncertainties on the eccentricities measurements. Black circles indicate the means for the entire dataset accounting for measurement errors.
                }
                \label{Fig1}
        \end{figure}

These 258 planets are distributed among the different multiplicities as described in Table \ref{table:1} and, even though more than half of the planets belongs to systems with $M$ = 2, all the considered multiplicity cases possess more than 10 planets, thus providing a statically relevant sample. Our dataset is plotted in Fig.~\ref{Fig1}, where we also show the planets of TRAPPIST-1 \citep{trappist} and of the solar system\footnote{https://nssdc.gsfc.nasa.gov/planetary/factsheet/}. This choice provides us with systems with 7 and 8 planets ($M$=7 and $M$=8) respectively, which we used to verify how our results extend to higher multiplicities. These two planetary systems, however, have not been included into our analysis and are shown only for comparison purposes.\\
        
        %-------------------------------------------------------------
        %                                             Simple A&A Table
        %-------------------------------------------------------------
        %
        \begin{table}
                \caption{Number of planets in our dataset divided by the multiplicity (M) of their host system. Also reported are the numbers of these planets that have been observed with the RV and transit method.}% title of Table
                \label{table:1}      % is used to refer this table in the text
                \centering                          % used for centering table
                \begin{tabular}{c c c c}        % centered columns (4 columns)
                        \hline\hline                 % inserts double horizontal lines
                        M & Total & RV & Tr \\    % table heading 
                        \hline                        % inserts single horizontal line
                        2 & 150 & 140 & 19 \\      % inserting body of the table
                        3 & 44 & 34  & 11 \\
                        4 & 26 & 21   & 12 \\
                        5 & 20 & 14  & 6 \\
                        6 & 18 & 18   & 8 \\ 
                        \hline                                   %inserts single line
                \end{tabular}
        \end{table}
        %
        
        %-------------------------------------------------------------
        %                 A figure as large as the width of the column
        %-------------------------------------------------------------
        
        \subsection{Multiplicity and eccentricity}
        
In analysing the dynamical properties of the exoplanets in our sample, instead of the median values of the eccentricity $e$ in the different multiplicity bins considered by \cite{limbach}, we focused on the relation between $M$ and the weighted average eccentricities. 

        The weights we used in computing the average values of each multiplicity bin were the inverse of the relative errors of the  eccentricities, so that
        
        \begin{equation}
        \overline{e}(M) =       \left[\sum{\left(\frac{e(M)_{i}}{\sqrt{\sigma_{e_{i}}^{2}}}\right)} \middle/        \sum{\left(\frac{1}{\sqrt{\sigma_{e_{i}}^{2}}}\right)}\right] \,
        ,\end{equation}
        
        where $\overline{e}(M)$ is the weighted average for a specific value of the multiplicity M (from 2 to 6), $e(M)_{i}$ are the individual eccentricity values of the exoplanets in the multiplicity bin, and $\sigma_{e_{i}}$ are the associated relative errors, which in the case of different values for positive and negative errors were computed as
        
        \begin{equation}
        \sigma_{e_{i}} = \frac{\sigma_{e_{i}}^{+} + \sigma_{e_{i}}^{-}}{2e_{i}} \,
        .\end{equation}
        
        \subsection{Multiplicity and dynamical state}

Our preliminary investigation of the dynamical state of the exoplanetary systems took advantage of the information supplied by the angular momentum deficit $AMD$ \citep{AMD}, which measures the difference between the angular momentum of a planetary system and the angular momentum the system would have if its planets were all on circular and planar orbits. 

As demonstrated by \cite{AMD2017}, the AMD remains stable after totally inelastic collisions, but increases when a body is ejected from the system. Therefore a system that experienced several ejections (hence decreasing in multiplicity $M$) should have a higher value of AMD with respect to a system that remained dynamically stable over all its history.
        
Following \cite{AMD}, the AMD is defined as
        
        \begin{equation}
        AMD = \sum_{k}{m_{k}\sqrt{Gm_{0}a_{k}}\left(1-\sqrt{1-e_{k}^2}cos(i_{k})\right)}
        ,\end{equation}  
        
where $G$ is the gravitational constant, $m_{k}$ the planetary masses (in Jovian masses), $m_{0}$ is the mass of the host star (in solar masses), and $a_{k}$, $e_{k}$, $i_{k}$ are the planetary semimajor axes (expressed in AU), eccentricities and inclinations, respectively. For every system, we considered as the reference inclination plane that of the orbit of the most massive planet and computed the relative inclinations of the other planets with respect to that orbit.
        
        When computing the AMD for the different exosystems in our sample we considered two datasets (see Table \ref{table:2}). The first  dataset (hereafter, the "large selection") comprises 10 exosystems (31 exoplanets) with known values of the mass, semimajor axis, eccentricity, and inclination for all planets. 
        
        The second one (hereafter the "strict selection") comprises 9 exosystems (28 exoplanets) selected with the additional requirement of also having known uncertainties on the four planetary parameters required for the computation of the AMD. To increase the statistics, in this case, we also considered those exosystems with $M>2$ for which inclination values were not available for at most one planet. 
        
        \begin{table}
                \caption{Number of systems at different multiplicities considered in the analysis of their dynamical state through the AMD.}             % title of Table
                \label{table:2}      % is used to refer this table in the text
                \centering                          % used for centering table
                \begin{tabular}{c c c c c}        % centered columns (4 columns)
                        \hline\hline                 % inserts double horizontal lines
%                       M & Large $E_{\sum}$ & Strict $E_{\sum}$ & 
                        M & Large $AMD$ & Strict $AMD$  \\    % table heading 
                        \hline                        % inserts single horizontal line
%                       2 & 7 & 5
                        2 & 5 & 3 \\      % inserting body of the table
%                       3 & 1 & 2 
                        3 & 1 & 2 \\
%                       4 & 3 & 3 
                        4 & 3 & 3 \\
%                       5 & 1 & 1 
                        5 & 0 & 0 \\
%                       6 & 1 & 1 
                        6 & 1 & 1 \\ 
                        \hline                                   %inserts single line
                \end{tabular}
        \end{table}
        
        \section{Results and discussion}
        
        \subsection{Multiplicity and eccentricity}
        
\begin{figure}
                \centering
                \includegraphics[width=\hsize]{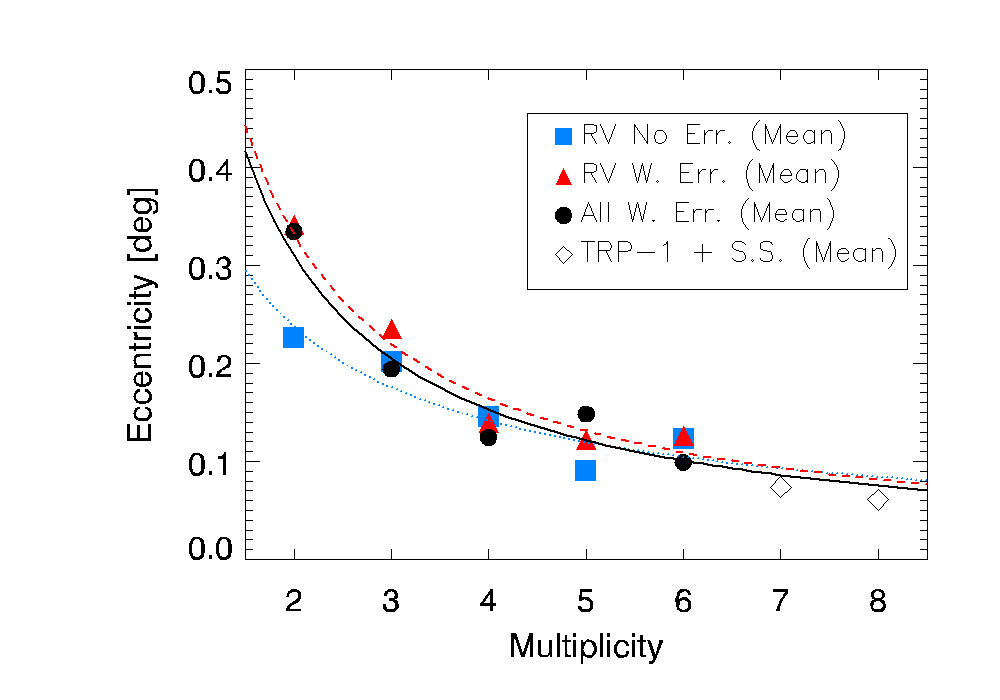}
                \caption{Power law fits. Blue squares indicate RV planets without considering errors, red triangles indicate RV planets with weights as errors, and black circles indicate all planets in the selected dataset with errors used as weights.
                }
                \label{Fig2}
        \end{figure}

        Fig.~\ref{Fig2} shows the $\overline{e}(M)$ behaviour, computed for three different subsets of our sample of exoplanets, and the associated power law fits.  Specifically, we computed the values of $\overline{e}(M)$ as (1) simple averages considering only the 227 exoplanets discovered with the RV method, i.e. similar to the original analysis by \cite{limbach}; (2) weighted averages as per Eq. 1 again considering only the 227 exoplanets discovered with the RV method; and (3) weighted averages as per Eq. 1 considering our entire dataset of 258 exoplanets.
        
        In Table \ref{table:3} we show the parameters of the fits and the associated Pearson correlation coefficients $R^2$. As shown by Table \ref{table:3}, there is a significant increase in the value of $R^2$ when measurement errors are used as weights in computing the averages. Another noticeable effect of the use of weighted averages is the increase in the value of the mean eccentricity for the case $M$ = 2. This results in the estimated power law to reproduce well all cases for $M>1$ instead of just $M>2$ as reported by \cite{limbach}.
        
        Finally, looking at Fig.~\ref{Fig2} one can see that the behaviour of the planets belonging to TRAPPIST-1 ($M=7$) and the solar system ($M=8$) is well reproduced by our fits, where that computed with the entire dataset most closely reproduces the average values associated with these two systems. Using the root-mean-squared deviation (RMSD) as an indicator of the goodness of the fits of the weighted averages for $M\geq7$, the case considering only RV data has RMSD = 0.020, while that considering the entire dataset has RMS = 0.013.%, resulting in:

        %-------------------------------------------------------------
        %                 A figure as large as the width of the column
        %-------------------------------------------------------------

        \begin{table}
                \caption{Fit parameters ($e(M)=AM^{B}$) and $R^{2}$ values for the case shown in Fig.~\ref{Fig2}.}             % title of Table
                \label{table:3}      % is used to refer this table in the text
                \centering                          % used for centering table
                \begin{tabular}{c c c c}        % centered columns (4 columns)
                        \hline\hline                 % inserts double horizontal lines
                        Case & A & B & $R^{2}$ \\    % table heading 
                        \hline                        % inserts single horizontal line
                        RV no error & 0.400 & -0.75 & 0.88\\      % inserting body of the table
                        RV & 0.668 & -1.01  & 0.96\\
                        Complete dataset & 0.630 & -1.02 & 0.95\\
                        \hline                                   %inserts single line
                \end{tabular}
        \end{table}

        \subsection{Multiplicity and dynamical state}

        %
        %-------------------------------------------------------------
        %                                             Two column Figure 
        %-------------------------------------------------------------
        \begin{figure}
                \centering
                \includegraphics[width=\hsize]{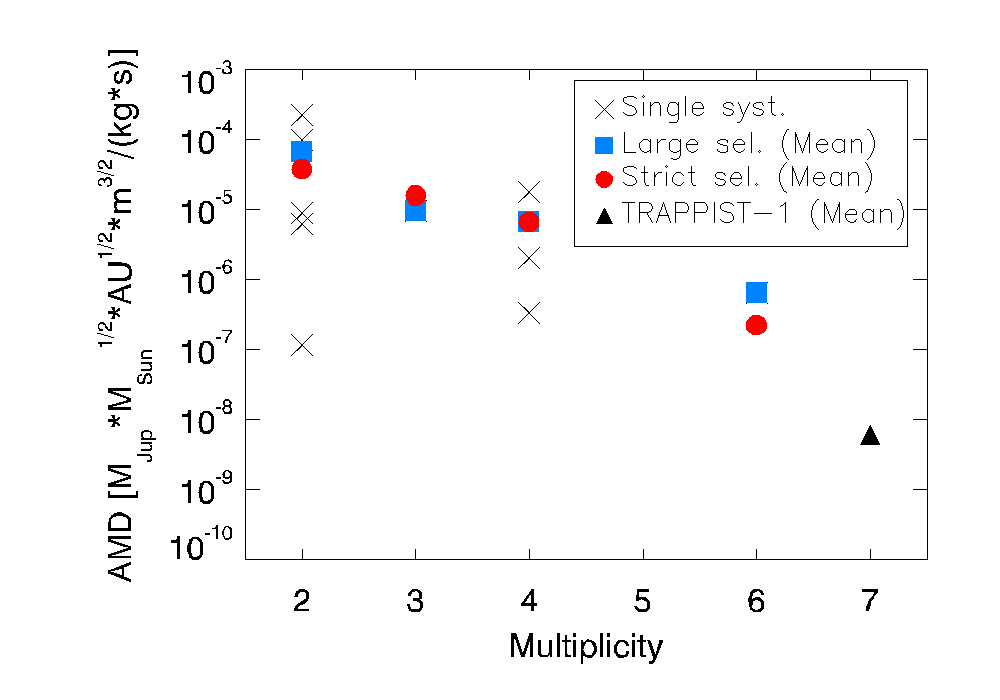}
        \caption{Angular momentum deficit of the systems considered. Crosses indicate values for the individual systems, blue squares indicate average values for the large selection sample, red circles indicate (with error bars) averages values for strict selection sample. The TRAPPIST-1 value is shown as a black triangle. Error bars are smaller than the symbol sizes.
        }
        \label{Fig3}
        \end{figure}

        Fig.~\ref{Fig3} shows the average values of the AMD of the exosystems in our large and strict selections for every multiplicity case.

Here we omitted the comparison with the solar system as it possesses masses and distances from the host star that are extremely different from those of the more compact exosystems in our sample. To this regard, \cite{Laskar97} also pointed out that the AMD of the solar system shows a bimodal behaviour with the inner system (i.e. the terrestrial planets) showing characteristics of an unstable system and the outer system (i.e the giant planets) more resembling a stable system.

        The analysis of the AMD shows an inverse trend with multiplicity for both the large and the strict selections. We did not attempt to fit the data as the number of systems analysed is statistically limited. Nevertheless, the fact that lower multiplicity systems display a higher value of the AMD, if confirmed when a larger sample of exosystems becomes available, could indicate that high-multiplicity systems are a common outcome of the planetary formation process, but that a significant portion of these systems form or evolves into unstable orbital configurations. 
        
        The loss of planets due to phases of dynamical instability by such systems (e.g. by planet-planet scattering events; see \citealt{weidenschilling,rasio}) would cause the increase of their AMD values and, consequently, of the eccentricities of their surviving planets. The resulting systems would then contaminate the population of exosystems with lower native multiplicity values, probably explaining the behaviours  observed in Fig. 2 and Fig. 3.
               
        \section{Conclusions} 
        
The results of this study confirm the reported anti-correlation between  the orbital eccentricity of exoplanets and the multiplicity of the host system they inhabit. The inclusion of the uncertainties on the measured planetary eccentricities in our analysis not only provides a more robust confirmation, but also allows for fitting with a single power law the average eccentricities for multiplicities comprised between $M=2$ and $M=6$. Furthermore, both the characteristics of TRAPPIST-1 ($M=7$) and of the solar system ($M=8$) are represented well by the same power law

\begin{equation}
e(M) = 0.630 M^{-1.02} \,
.\end{equation}
                
In the attempt to provide a physical explanation of the reported behaviour, we made a preliminary analysis of the correlation between the angular momentum deficit AMD of the systems included in our sample and their multiplicity values. While the data are not sufficient to provide a conclusive answer, we observed an anti-correlation between the multiplicity of the exosystems and their AMD as well. Should this AMD-multiplicity anti-correlation of exosystems be verified once larger and more precise datasets become available, it would provide us with a deeper insight both on the planetary formation process and on the place of the solar system in an exoplanetary context.\\
Therefore this could be an interesting field of research when the next generation of space telescopes dedicated to exoplanetary systems is ready, starting with the NASA Transiting Exoplanet Survey Satellite (TESS; \citealt{Ricker_TESS}), which is planned to be operative in the next year.
 
Specifically, our results raise the possibility that high-multiplicity systems such as the solar system are a relatively common outcome of the planetary formation process. Our results also suggest that, unlike the solar system, a possibly large portion of these systems form or evolve into dynamically unstable orbital configurations. The ejection of one or more planets from these unstable systems would naturally explain our results by decreasing the multiplicities of the systems increasing their AMD and eccentricities and/or inclinations of their surviving planets at the same time. As a consequence, the population of low-multiplicity exosystems we observe is likely to be contaminated by former high-multiplicity systems that became dynamically unstable and lost some of their planets during their histories.

        
        \bibliographystyle{aa}

\end{document}